\documentclass[sigconf]{acmart}
\usepackage{xspace}
\usepackage{longtable}
\usepackage{array}
\usepackage[table]{xcolor}

\usepackage{color}
\usepackage{nameref}
\usepackage{multirow}
\usepackage{tabularx} 
\usepackage{censor}
\usepackage{nameref}
\usepackage{enumitem}

\usepackage[most]{tcolorbox}

\usepackage{xcolor}          

\definecolor{lightblue}{RGB}{0, 0, 100}

\newtcolorbox{MyBox}{
  colback=white,
  colframe=lightblue,
  fonttitle=\bfseries,
  coltitle=black,
  sharp corners,
  boxrule=1pt,
  left=5pt,
  right=5pt,
  top=5pt,
  bottom=5pt,
  breakable
}

\AtBeginDocument{%
  }

\setcopyright{acmlicensed}
\copyrightyear{2018}
\acmYear{2018}
\acmDOI{XXXXXXX.XXXXXXX}
\acmConference[Boatse '26]{}{June 03--05,
  2018}{Woodstock, NY}
\acmISBN{978-1-4503-XXXX-X/2018/06}

\begin{document}

\title{Bita: A Conversational Assistant for Fairness Testing}

\author{Keeryn Johnson}
\email{keeryn.johnson@ucalgary.ca}
\orcid{}
\affiliation{%
  \institution{University of Calgary}
  \city{Calgary}
  \state{Alberta}
  \country{Canada}}

\author{Cleyton Magalhaes}
\email{cleyton.vanut@ufrpe.br}
\orcid{}
\affiliation{%
  \institution{UFRPE}
  \city{Recife}
  \state{Pernambuco}
  \country{Brazil}}

\author{Ronnie de Souza Santos}
\email{ronnie.desouzasantos@ucalgary.ca}
\orcid{0000-0003-3235-6530}
\affiliation{%
  \institution{University of Calgary}
  \city{Calgary}
  \state{Alberta}
  \country{Canada}
  }

\renewcommand{\shortauthors}{Johnson et al.}
\begin{abstract}
Bias in AI systems can lead to unfair and discriminatory outcomes, especially when left untested before deployment. Although fairness testing aims to identify and mitigate such bias, existing tools are often difficult to use, requiring advanced expertise and offering limited support for real-world workflows. To address this, we introduce Bita, a conversational assistant designed to help software testers detect potential sources of bias, evaluate test plans through a fairness lens, and generate fairness-oriented exploratory testing charters. Bita integrates a large language model with retrieval-augmented generation, grounding its responses in curated fairness literature. Our validation demonstrates how Bita supports fairness testing tasks on real-world AI systems, providing structured, reproducible evidence of its utility. In summary, our work contributes a practical tool that operationalizes fairness testing in a way that is accessible, systematic, and directly applicable to industrial practice.
\end{abstract}

\ccsdesc[500]{Software and its engineering~Software creation and management~Software development process management}
\keywords{fairness testing, software engineering bot, testing tool}



\maketitle

\section{Introduction}
\label{sec:introduction}
Fairness testing is a growing area of concern in the development of artificial intelligence (AI) systems, aiming to identify and mitigate discriminatory behaviour in software \cite{chen2024surveyanalysis, galhotra2017fairness, brun2018software}. As AI becomes increasingly influential in decisions in various contexts, from hiring, finance, and criminal justice \cite{chen2024surveyanalysis, galhotra2017fairness}, it can significantly impact individuals and communities. Ensuring that these systems operate fairly has become both an ethical requirement and a technical challenge \cite{chen2024surveyanalysis, galhotra2017fairness, brun2018software, santos2025debt}. In this context, fairness testing plays a key role in ensuring that algorithms produce equitable outcomes across demographic groups, such as race, gender, or socioeconomic status \cite{galhotra2017fairness}. Unlike traditional testing, fairness testing emphasizes the ethical and social dimensions of software behaviour, focusing on instances where the model may treat individuals or groups unfairly \cite{sun2024translation, chen2024surveyanalysis}. Hence, it extends the scope of conventional testing, emphasizing not just correctness and reliability, but also justice and accountability in software behavior \cite{chen2024surveyanalysis}.

Despite its growing importance, fairness testing still faces several obstacles that challenge its adoption in software industry practice. One major issue is the lack of accessible, reliable, and adaptable tools to support fairness evaluations. Many existing tools are research prototypes that require advanced machine learning expertise, offer limited documentation, and are difficult to integrate into real-world development workflows \cite{nguyen2024literature}. This situation is further intensified by the diversity and inconsistency of fairness definitions, which are often context dependent and sometimes mutually incompatible \cite{holstein2019improving}. As a result, professionals face difficulties in selecting appropriate fairness criteria and designing effective tests that align with their system goals \cite{chen2024surveyanalysis}. In addition, fairness concerns are often rooted in data and model behavior, demanding machine learning knowledge and analytical capacity that many testing professionals lack \cite{nguyen2024literature, santos2025software}. Resource constraints further exacerbate the problem, as fairness testing can be time-consuming and computationally intensive, posing a challenge in fast-paced, delivery-driven environments \cite{aggarwal2019blackbox, chen2024surveyanalysis}.

Considering these persistent barriers, there remains a need for fairness testing support that is both accessible and adaptable to the realities of software practice \cite{santos2025software, nguyen2024literature}. Existing tools, though methodologically sound, often remain detached from testers’ day-to-day workflows and assume technical expertise beyond that of many professionals. To respond to this gap, we introduce \textit{Bita}, a conversational assistant that helps software testers identify and address fairness issues during AI system validation. Unlike prior tools that focus narrowly on algorithmic metrics or mitigation, Bita reframes fairness testing as a tester-centric activity. Through natural language interaction, it helps testers (i) identify potential fairness concerns in systems, (ii) evaluate test plans from a fairness perspective, and (iii) generate exploratory testing charters that incorporate fairness considerations.

Overall, the contributions of our paper are threefold. First, we introduce Bita, a conversational assistant that integrates large language models (LLMs) and retrieval augmented generation to support fairness testing activities. Second, we present an evaluation of Bita through illustrative workloads, including bias identification, test plan review, and exploratory charter generation on real-world AI systems, together with a structured comparison between Bita and existing fairness testing tools. Third, we offer a practice-oriented contribution by framing fairness testing as an activity that can be embedded in software validation workflows, making fairness considerations accessible to practitioners beyond specialized ML experts.

From this introduction, this paper is organized as follows. Section~\ref{sec:background} expands the discussion on fairness and its societal impacts, reviews existing fairness testing tools, and explores the use of bot-based assistants in software development. Section~\ref{sec:method} describes our methodology. Section~\ref{sec:bita} and ~\ref{sec:findings} introduce our proposed tool, followed by a discussion in Section~\ref{sec:discussion}. Finally, Section~\ref{sec:conclusions} summarizes our contributions and presents directions for future work.
\section{Background}
\label{sec:background}

This section synthesizes relevant concepts in software fairness, fairness testing tools, and the use of bots in software engineering to contextualize the development of \textit{Bita}.

\subsection{Software Fairness and Algorithmic Bias}
Software fairness concerns the principle that systems should produce equitable outcomes for all users, independent of sensitive attributes such as age, race, or gender \cite{soremekun2022softwarefairnessanalysissurvey, brun2018software}. In practice, similar inputs should lead to comparable outcomes regardless of the user’s identity. Achieving fairness depends on both the technical soundness of models and data, and on an understanding of the social contexts in which systems operate \cite{holstein2019improving, santos2025debt}. When fairness concerns are neglected, systems tend to accumulate what has been described as \textit{fairness debt} \cite{santos2025debt}, in which unaddressed bias compounds over time and becomes increasingly difficult to resolve. In this context, algorithmic bias emerges when machine learning models reproduce or amplify inequalities that exist in historical or unbalanced data \cite{brun2018software, kordzadeh2022algorithmic, hooker2021moving}. Bias may result from skewed datasets, inappropriate feature selection, or unintended correlations between sensitive and non-sensitive variables \cite{li2022trainingdata, chen2024surveyanalysis, mehrabi2021survey, dehal2024discrimination}. These flaws can lead to fairness bugs \cite{chen2024surveyanalysis} that systematically disadvantage certain subgroups, particularly in domains such as hiring, credit scoring, policing, or healthcare \cite{dehal2024discrimination, fountain2022moon, zliobaite2017measuring, kleinberg2018discrimination}. Consequences of bias extend beyond technical failure and affect social trust. When decisions produced by AI systems are perceived as discriminatory, public confidence erodes and demands for oversight increase \cite{glikson2020human, ferrara2023howfararewe, santoni2021four, smuha2021race}. Organizations also face legal and reputational risks when their systems are shown to generate unequal outcomes \cite{daloisio2025fairweconceptualizationautomated}. These challenges underscore the need for systematic fairness testing to detect and mitigate bias before deployment \cite{chen2024surveyanalysis}.

\subsection{Fairness Testing Tools}
Fairness testing refers to the systematic evaluation of software or AI systems to determine whether their behaviour produces unequal outcomes across demographic groups \cite{chen2024surveyanalysis, brun2018software}. This form of testing extends traditional software testing by emphasizing equity and accountability in addition to correctness and reliability \cite{soremekun2022softwarefairnessanalysissurvey, santos2025debt}. Fairness testing helps identify and mitigate algorithmic bias early in development, thereby reducing the risk of embedding discriminatory effects in deployed systems \cite{holstein2019improving, chen2024surveyanalysis}. A broad range of tools has been developed to support fairness testing \cite{chen2024surveyanalysis}. Some focus on bias detection and reporting (e.g., \textit{FairTest}, \textit{Themis}, \textit{Aequitas}), others address data analysis and preprocessing (e.g., \textit{LTDD}, \textit{FairMask}, \textit{Fairway}), and others aim at model repair or fairness optimization (e.g., \textit{FairRepair}, \textit{Parfait-ML}, \textit{Fairea}). Integrated frameworks such as \textit{IBM AIF360}, \textit{Scikit-fairness}, and \textit{LiFT} cover multiple stages of the machine learning pipeline, while domain-specific tools like \textit{DeepInspect} and \textit{BiasAmp} (image classification), \textit{CheckList}, \textit{ASTRAEA}, and \textit{DialogueFairness} (NLP), as well as \textit{BiasFinder} and \textit{BiasRV} (sentiment analysis), address targeted applications. Despite their range, the adoption of these tools remains limited. Evaluations show frequent usability issues such as broken dependencies, limited documentation, and assumptions of high machine learning expertise \cite{nguyen2024literature}. Only a small fraction of tools meet minimal usability standards for integration into practical testing workflows. Furthermore, most tools are restricted to narrow problem types (e.g., binary classification) and rely heavily on benchmark datasets. The absence of user-centered design and workflow compatibility continues to limit their applicability in industrial settings \cite{nguyen2024literature, santos2025software}.

\subsection{Bots in Software Engineering}
Bots in software engineering are automated agents that assist developers and testers by monitoring, analyzing, and responding to changes in development environments \cite{santhanam2022botsinse, lambiase2024motivations}. These agents contribute to software projects by automating repetitive or time-consuming tasks and by facilitating coordination across distributed teams \cite{erlenhov2020characteristicsofbots}. Their use has expanded across activities such as code review, issue management, continuous integration, and testing, where they reduce manual effort and enhance process consistency \cite{santhanam2022botsinse}. In addition to task automation, bots support communication and knowledge sharing by posting reminders, summaries, and updates in collaborative environments such as GitHub and Slack \cite{erlenhov2020characteristicsofbots}. They also assist during debugging and testing by providing on-demand guidance, contextual feedback, and documentation references \cite{lambiase2024motivations}. These functions contribute to continuous quality assurance and allow human developers to concentrate on higher-level reasoning and problem solving \cite{wessel2022benefits, shihab2022futureofbots}. Research on bots demonstrates their capacity to mediate collaboration and to sustain productivity across software teams~\cite{wessel2022benefits, santhanam2022botsinse, erlenhov2020characteristicsofbots}. Their integration into testing workflows can be a promising direction for supporting complex and socially sensitive activities such as fairness testing. Conversational bots, in particular, can translate specialized knowledge into accessible, context-aware dialogue ~\cite{santhanam2022botsinse, wessel2022benefits}, which would make fairness analysis more approachable for practitioners who lack advanced expertise in machine learning or ethics.

\section{Method} \label{sec:method}

This study follows the engineering research method, also known as Design Science~\cite{alph2021empiricalstandardssoftwareengineering}, as the primary contribution is a software artifact developed to address a practical challenge: assisting software testers in identifying and mitigating algorithmic bias during AI system validation. The artifact, named \textit{Bita}, was conceived as a conversational bot~\cite{santhanam2022botsinse, wessel2022benefits, erlenhov2020characteristicsofbots}, integrating recent advances in human–bot interaction to deliver fairness testing support through accessible, dialogue-based exchanges grounded in established fairness literature~\cite{chen2024surveyanalysis, brun2018software}.

\subsection{Tool Development}
The development of \textit{Bita} was guided by three design principles focused on usability, adaptability, and reliability for integration into existing testing workflows. The first principle concerns usability through natural language interaction. Conversational interfaces have proven effective in supporting software engineering activities, including code review, quality assurance, and team coordination~\cite{santhanam2022botsinse, erlenhov2020characteristicsofbots, lambiase2024motivations}. These systems reduce the cognitive effort required to use technical tools and make complex processes more approachable. Following this rationale, \textit{Bita} was implemented as a conversational bot that delivers fairness testing guidance through dialogue. Instead of configuring fairness libraries or writing scripts, testers communicate directly with the assistant in natural language to describe their systems and testing goals.

The second principle focuses on adaptability. A large language model forms the core of \textit{Bita}, allowing it to interpret incomplete or ambiguous tester input while maintaining contextual awareness~\cite{fan2023large, lin2025can}. This capability supports flexible reasoning across varied testing contexts. The conversation management component preserves dialogue history, ensuring that responses remain coherent across multiple interactions. Prompt design was informed by recent research on prompt engineering~\cite{hewing2024promptcanvasliteraturebasedpractitioner}, aligning fairness-related prompts with recognized software testing practices. As a result, \textit{Bita} produces tailored recommendations that help testers reason about fairness risks and design test cases reflecting the system’s operational context.

The third principle emphasizes reliability through RAG~\cite{gao2023retrieval, dong2025understand}. The backend architecture integrates an information retrieval layer connected to a curated corpus of fairness testing research \cite{chen2024surveyanalysis, patel2022combofairnesstesting, yin2024blackboxfairness, xie2020deepreinforcementfairness, monteiro2024fairnessevalconditionaltesting, mamman2023searchbasedfairnesstestingoverview}, tool documentation, and empirical studies~\cite{santos2025software, nguyen2024literature}. Retrieved outputs are combined with user prompts to ground responses in verified knowledge rather than relying solely on generative inference. This hybrid design increases transparency and supports traceability by linking each suggestion to relevant sources. Such grounding is particularly important in conversational bots for software engineering, where accuracy and verifiability are required~\cite{wessel2022benefits, shihab2022futureofbots, santhanam2022botsinse, erlenhov2020characteristicsofbots}.

\subsection{Task Definition}
To structure \textit{Bita}’s development, three representative tasks were defined, each corresponding to a typical activity in fairness testing practice. The first task, bias identification, enables the assistant to detect potential fairness concerns such as sensitive attributes, biased relationships, and overlooked subgroups within a system description. The second task, test plan evaluation, allows the assistant to review existing test plans from a fairness perspective, identifying missing cases and weaknesses in demographic coverage. The third task, exploratory charter generation, involves the assistant proposing practical charters that guide testers in examining fairness-related behaviours during exploratory testing sessions. These tasks represent the central stages of fairness testing supported by the assistant: planning, evaluation, and execution.

\subsection{Validation}
Our evaluation of Bita combined conceptual and empirical procedures, following two established empirical research guidelines~\cite{alph2021empiricalstandardssoftwareengineering, hasselbring2021benchmarking}. A comparative analysis positioned Bita relative to existing fairness testing tools~\cite{chen2024surveyanalysis, nguyen2024literature}, with attention to usability, adaptability, and integration dimensions. The assistant was subsequently applied in simulated scenarios derived from two real-world AI systems~\cite{mari2020libras, botocario} to assess its ability to identify fairness risks, propose mitigation strategies, and generate exploratory testing charters. This multi-stage validation was designed to demonstrate how Bita facilitates fairness testing activities in realistic development settings.
\section{Bita: Bias Identification and Testing Assistant} \label{sec:bita}
Bita (\textit{Bias Identification and Testing Assistant}) is a conversational system developed to support fairness testing in AI applications through natural language interaction. The tool follows a modular architecture organized into frontend, backend, and database layers. The frontend offers a web-based interface through which testers describe the system under evaluation, upload test plans, and request exploratory testing guidance. The backend integrates an RAG pipeline that combines LLM reasoning with document retrieval from a curated corpus of fairness research and tool documentation. The database stores session history, retrieved evidence, and user feedback, ensuring continuity across interactions and enabling the refinement of subsequent testing sessions.

The interaction begins when testers describe the system under test, specifying its purpose, inputs, outputs, and target users. Based on this description, Bita provides three forms of support aligned with common fairness testing activities. The \textbf{Bias Detection} feature analyzes the system under test description to list potential bias sources, such as neglected subgroups, imbalanced data attributes, or fairness bugs that may emerge during system use. The \textbf{Plan Check} component processes test plans to identify missing demographic coverage, untested scenarios, or hidden correlations between sensitive and non-sensitive attributes. The \textbf{Exploratory Charter Generation} component produces structured testing prompts that guide practitioners in investigating fairness risks in realistic contexts. These features were built to address the planning, evaluation, and execution stages of fairness testing, promoting more reflective and systematic practices among testers. 

Prompt construction in Bita is grounded in a retrieval-enhanced workflow. User queries are vectorized through a SentenceTransformer encoder and matched against a fairness literature index that includes sources such as fairness surveys, tool evaluations, and methodological guidelines \cite{chen2024surveyanalysis, patel2022combofairnesstesting, yin2024blackboxfairness, xie2020deepreinforcementfairness, monteiro2024fairnessevalconditionaltesting, mamman2023searchbasedfairnesstestingoverview}. Retrieved document segments are combined with the user query and context to form structured prompts, which are then processed by either OpenAI’s ChatGPT or Google’s Gemini models, dynamically selected at runtime. The system employs prompt-engineering strategies such as few-shot prompting \cite{brown2020language}, instruction-based prompting \cite{wei2022finetuned}, and role prompting \cite{zhou2023survey} to maintain contextual accuracy and minimize hallucinations (example below). Guardrails embedded in the RAG pipeline ensure that responses remain within the scope of fairness testing and are supported by retrieved evidence.

\begin{MyBox} \footnotesize
"You are a software fairness assistant that explains fairness testing concepts in clear and practical terms. The user previously asked about fairness testing in AI systems. The retrieved documents describe that demographic parity is a group-based fairness metric, while equalized odds ensures parity in true and false positive rates across subgroups. Using this context, explain the difference between demographic parity and equalized odds, and when each should be applied in fairness testing."
\end{MyBox}

\begin{figure*}[h]
\centering
\includegraphics[width=1\linewidth]{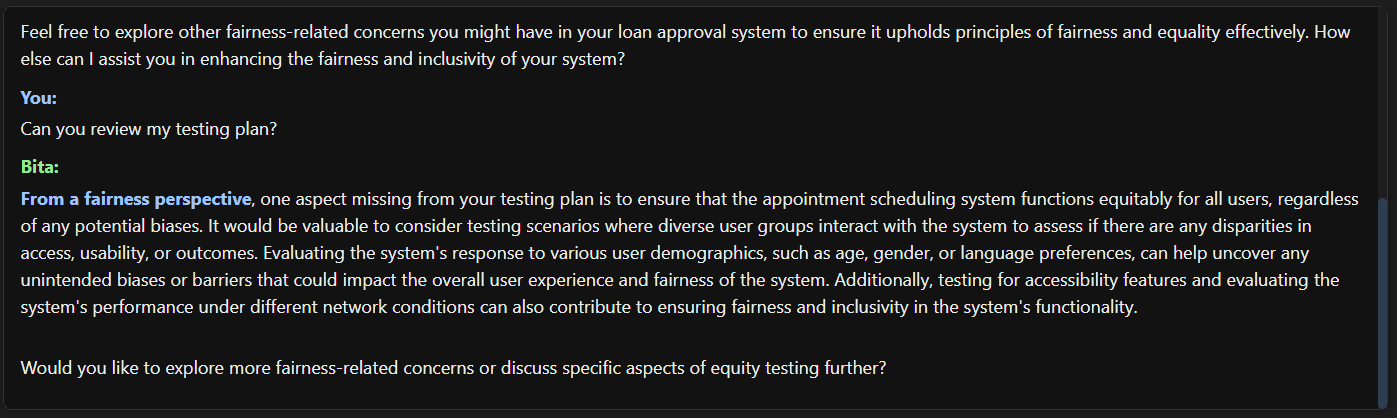}
\caption{Bita’s Interactive Interface in Use}
\label{fig:enter-label}
\end{figure*}

Responses are returned to the frontend and stored in an SQL database together with session data such as user queries, system descriptions, and generated outputs. This design enables continuity across multiple interactions and supports iterative refinement of testing guidance. As users provide additional context, subsequent prompts incorporate prior information, allowing Bita to deliver progressively more specific and contextually relevant support. All interaction data, including user inputs, retrieved literature, and generated responses, are stored in an SQL database. This structure supports multi-turn dialogue continuity, reproducibility of previous sessions, and incremental knowledge accumulation for each user. This architecture allows testers to refine their fairness assessments progressively, integrating updated system descriptions, test plans, and outcomes over time. Figure~\ref{fig:enter-label} illustrates the overall system interface, while the online repository provides implementation details: \url{https://figshare.com/s/b71e5093042a151a13f3}
. The live version of Bita is available at \url{https://bitatesting.ca/}
.

\section{Results} 
\label{sec:findings}

Our results focused on exploring how Bita performs on fairness testing tasks and how it compares with existing fairness testing tools.

\subsection{Performing Tasks}

Our results demonstrate how Bita operationalizes the three design principles—usability, adaptability, and reliability—through the tasks defined in Section~\ref{sec:method}: bias identification, plan evaluation, and exploratory charter generation. Each task was applied to two real-world AI systems: a sign language translation tool~\cite{mari2020libras} and a Smart Lipstick application~\cite{botocario}.

\begin{itemize}
\item \textbf{Bias Identification.} In the sign language translator~\cite{mari2020libras}, Bita identified fairness concerns related to regional dialect coverage, diversity of signers, and accessibility for users with distinct physical abilities. In the Smart Lipstick application~\cite{botocario}, it emphasized the need for testing across different skin tones, handling facial asymmetries, and ensuring usability for individuals with limited motor control. These examples show Bita’s \textit{adaptability}, as it adjusts fairness reasoning to the context of each system rather than producing generic fairness statements. They also reflect \textit{reliability}, since recommendations are grounded in documented fairness concepts retrieved from the literature.
\end{itemize}

\begin{itemize}
\item \textbf{Plan Evaluation.} Test plans created for each system were analyzed by Bita, which identified overlooked demographic subgroups, edge cases in input variability, and correlations between sensitive and non-sensitive attributes. In the sign language translator, Bita noted the absence of cases involving older signers or variations in signing speed. In the Smart Lipstick application, it highlighted untested scenarios across lighting conditions and camera resolutions that might influence fairness outcomes. These findings illustrate \textit{usability}, as testers could obtain structured feedback through natural language interaction without the need to configure fairness libraries. They also demonstrate \textit{workflow integration}, since the assistant complements existing testing documentation and practices rather than replacing them.
\end{itemize}

\begin{itemize}
\item \textbf{Exploratory Testing Charter Generation.} For each system, Bita generated exploratory testing charters that provided structured prompts to guide fairness-oriented investigation. The resulting charters support \textit{usability} by translating fairness concepts into practical testing actions, and \textit{adaptability} by aligning recommendations with each system’s functionality and context of use. These charters can be applied directly by practitioners to plan and conduct fairness-aware exploratory testing sessions.
\end{itemize}

Overall, these results indicate that Bita provides accessible, context-sensitive, and traceable support for fairness testing. The combination of conversational interaction, retrieval-based reasoning, and structured task execution demonstrates its potential to integrate fairness testing into the everyday practice of software validation.

\subsection{Comparison with Existing Tools}

We compared Bita against 42 fairness testing tools reported in the literature~\cite{chen2024surveyanalysis}. Table \ref{tab:full-comparison} summarizes this comparison in six dimensions of evaluation: support of fairness testing, bias detection, evaluation of tests, direct testing usage, conversational interaction, and grounding in the fairness literature. These dimensions move beyond algorithmic performance to also consider usability, accessibility, and integration into real-world testing workflows.  

\begin{table*}[t]
\centering
\caption{Comparison of Bita with existing fairness testing tools}
\label{tab:full-comparison}
\footnotesize
\begin{tabular}{p{1.8cm} p{2.5cm} p{1.5cm} p{1.5cm} p{1.5cm} p{1.5cm} p{1.5cm} p{1.5cm}}
\toprule
\textbf{Tool} & \textbf{Application} & \textbf{Component} & \textbf{Support to Fairness Testing} & \textbf{Bias Detection} & \textbf{Test Evaluation} & \textbf{Direct Testing Usage} & \textbf{Conversational Interface with RAG Grounding} \\
\midrule
FairTest & General ML & Model & Yes & Yes & No & No & No \\
Themis & Classification & Model & Yes & Yes & No & No & No \\
Aequitas & Classification & Model & Yes & Yes & No & No & No \\
ExpGA & Tabular data, text & Model & Yes & No & No & No & No \\
fairCheck & Tabular data & Model & Yes & Yes & No & No & No \\
MLCheck & Classification & Model & Yes & Yes & No & No & No \\
LTDD & Tabular data & Data & Yes & No & No & No & No \\
Fair-SMOTE & Tabular data & Model & Yes & No & No & No & No \\
FairMask & Tabular data & Data & Yes & No & No & No & No \\
Fairway & Tabular data & Data & Yes & No & No & No & No \\
Parfait-ML & Tabular data & ML program & Yes & No & No & No & No \\
Fairea & Benchmarking & Model & Yes & No & No & No & No \\
IBM AIF360 & Tabular data & Platform & Yes & No & No & No & No \\
I\&D & Classification & Model & Yes & No & No & No & No \\
scikit-fairness & Tabular data & Platform & Yes & No & No & No & No \\
LiFT & Tabular data & Platform & Yes & No & No & No & No \\
FairVis & Tabular data & Model & Yes & Yes & No & No & No \\
BiasAmp & Image classifier & Model & Yes & No & No & No & No \\
MAAT & Tabular data & Data & Yes & Yes & No & No & No \\
FairEnsembles & Tabular data & ML program & Yes & No & No & No & No \\
FairRepair & Tabular data & Model & Yes & No & No & No & No \\
SBFT & Tabular data & Model & Yes & Yes & No & No & No \\
ADF & Tabular data & Model & Yes & Yes & No & No & No \\
EIDIG & Tabular data & Model & Yes & Yes & No & No & No \\
NeuronFair & Tabular data, face images & Model & Yes & Yes & No & No & No \\
DeepInspect & Image & Model & Yes & Yes & No & No & No \\
CMA & Language models & Model & Yes & No & No & No & No \\
FairNeuron & Tabular data & Model & Yes & Yes & No & No & No \\
RULER & Tabular data & Model & Yes & No & No & No & No \\
TestSGD & Tabular data, text & Model & Yes & Yes & No & No & No \\
DICE & Tabular data & Model & Yes & Yes & No & No & No \\
ASTRAEA & NLP & Model & Yes & Yes & No & No & No \\
MT-NLP & NLP & Model & Yes & No & No & No & No \\
BiasFinder & Sentiment analysis & Model & Yes & Yes & No & No & No \\
BiasRV & Sentiment analysis & Model & Yes & Yes & No & No & No \\
NERGenderBias & NER & Model & Yes & Yes & No & No & No \\
CheckList & NLP & Model & Yes & Yes & No & No & No \\
DialogueFairness & Conversational AI & Model & Yes & Yes & No & No & No \\
BiasAsker & Conversational AI & Model & Yes & Yes & No & No & No \\
REVISE & CV datasets & Data & Yes & No & No & No & No \\
AequeVox & Speech recognition & Model & Yes & No & No & No & No \\
\textbf{Bita (ours)} & General AI Testing & System-level Assistant & Yes & Yes & Yes & Yes & Yes \\
\bottomrule
\end{tabular}
\end{table*}

The comparison of 42 tools shows clear imbalances across these six dimensions. All surveyed tools (100\%) claim to support fairness testing, but in most cases this support is restricted to model-level checks. Bias detection is the most mature dimension, with 39 tools (93\%) offering group comparisons or fairness metrics. In contrast, only 14 tools (33\%) extend their functionality to the evaluation of tests, and even these focus mainly on classification or tabular data contexts. Beyond these two dimensions, most tools remain difficult to integrate into everyday practice, as they are released as research prototypes in the form of libraries or pipelines that assume advanced ML expertise and require manual configuration. This indicates that while fairness testing support and bias detection are relatively well established, the other dimensions remain underdeveloped.  

The gaps become particularly evident in the remaining three dimensions. None of the 42 surveyed tools provide direct testing usage, such as exploratory charters that could embed fairness concerns into iterative, human-driven activities. Similarly, none of the tools support conversational interaction, and none ground their outputs in fairness literature through retrieval-augmented methods. These missing dimensions are critical for usability, accessibility, and transparency, yet they are entirely absent from the current landscape. In contrast, Bita spans all six criteria: it provides broad support for fairness testing, incorporates bias detection, extends to test evaluation, enables direct testing usage through fairness-oriented charters, and lowers barriers for testers via a conversational interface with responses grounded in curated research. This positions Bita as a practical assistant offering structured tasks and evaluation criteria for systematic and reproducible assessment of fairness testing assistants.

\section{Discussion}
\label{sec:discussion}

This section examines the results in relation to prior studies and summarizes their implications for both research and practice.

\subsection{Comparison to Literature and Novelty}
Previous studies show that most fairness testing tools concentrate on bias detection, presuppose substantial ML expertise, and remain difficult to integrate into industrial workflows~\cite{chen2024surveyanalysis, nguyen2024literature}. Bita responds to these limitations by introducing a conversational assistant that operationalizes fairness testing through interactive dialogue, structured guidance, and traceable recommendations. Research on software engineering bots has shown their value in automating complex tasks, facilitating collaboration, and mediating developer knowledge~\cite{santhanam2022botsinse, wessel2022benefits, erlenhov2020characteristicsofbots}. Extending this perspective, Bita demonstrates how conversational agents can support socially oriented testing activities, translating fairness concepts into actionable steps. Unlike algorithm-centric tools~\cite{chen2024surveyanalysis, nguyen2024literature}, Bita incorporates socio-technical reasoning~\cite{santos2025debt} by generating exploratory charters, contextual test evaluations, and fairness-aware suggestions that align with testers’ established practices. Our results indicate that structured conversational tasks can be used to evaluate how bots contribute to fairness testing in a systematic and reproducible manner. The proposed tasks define early baselines for assessing such assistants, while the identified dimensions foreground criteria that have received little attention in prior work, such as usability, adaptability, and workflow alignment. Bita thus extends fairness testing beyond technical performance and into the domain of human–AI collaboration, showing how conversational bots can make fairness testing accessible to practitioners who lack specialized expertise. Through natural language interaction and integration with exploratory testing workflows, Bita demonstrates a new form of bot-mediated assistance for fairness evaluation.

\subsection{Implications for Research and Practice}
Our study has implications for both research and professional practice. For research, this work offers a foundation for exploring how conversational assistants can support fairness testing in software engineering. Our study advances understanding of how usability, adaptability, and workflow integration can be incorporated into fairness-oriented tools. Bita represents an initial version of this approach, and future research will extend its capabilities, validate its effectiveness in additional contexts, and refine its design through iterative improvement. For practice, Bita provides a workflow-integrated approach that addresses barriers frequently identified in the literature~\cite{santos2025software, nguyen2024literature}. The conversational interface enables testers without ML expertise to identify fairness concerns, evaluate test plans, and generate exploratory charters. The retrieval-augmented architecture ensures that recommendations remain transparent and supported by recognized sources. Therefore, Bita's first version establishes the foundation for a continuous improvement process aimed at expanding features, refining dialogue capabilities, and deepening alignment with professional testing practices. Through this evolution, Bita is expected to become a more comprehensive and dependable assistant for fairness testing in real-world projects.

\subsection{Threats to Validity}
Following established engineering research standards~\cite{alph2021empiricalstandardssoftwareengineering}, some validity threats must be acknowledged. Bita was evaluated in a controlled environment using a fixed set of user prompts, which may not reflect the range of interactions that occur in real testing situations. Assumptions about tester experience and input formulation were not independently verified and may have influenced the results. The evaluation dimensions of usability, adaptability, and integration were derived from the literature but have not yet been externally confirmed, so the findings should be regarded as preliminary. Further assessments with professional testers are necessary to examine how the assistant performs in practice and to refine the evaluation framework. The analysis was limited to two AI systems, which constrains generalization to other domains and forms of fairness testing. Overall, future evaluations are necessary to verify and consolidate the results.

\section{Conclusions and Future Work} \label{sec:conclusions}
This paper introduced \textit{Bita}, a conversational bot designed to assist software testers in recognizing and addressing fairness concerns in AI systems. Through natural language interaction and retrieval-augmented grounding in fairness literature, Bita acts as an interactive collaborator that guides testers in reasoning about fairness during validation activities. The assistant lowers the expertise required to incorporate fairness considerations into software testing workflows. Our preliminary results indicate that Bita provides contextually appropriate recommendations and practical guidance to testers. Compared with existing tools, Bita demonstrates stronger usability, transparency, and workflow alignment, responding to current challenges observed in fairness testing practices. Our future work will continue the iterative development of the bot by conducting in-situ studies with practitioners, analyzing adoption barriers, and extending its reasoning capabilities toward adaptive, context-sensitive support for fairness testing in software engineering.

\section*{Acknowledgments}
This work was supported by the Natural Sciences and Engineering Research Council of Canada (NSERC), Discovery Grant RGPIN-2024-06260, and by Alberta Innovates through the Advance Program, Project Number 24506125.

\bibliographystyle{ACM-Reference-Format}
\bibliography{bibliography}

\appendix

\end{document}